\documentclass[11pt,a4paper]{article}

\usepackage{graphicx}
\graphicspath{{figures/}}
\usepackage{amsmath,amsfonts,amsthm}

\numberwithin{equation}{section}

\newcommand{\beq}{\begin{equation}}
\newcommand{\eeq}{\end{equation}}
\newcommand{\Lc}{\mathcal{L}}
\newcommand{\ds}{\displaystyle}

\newtheorem{thm}{Theorem}
\newtheorem{prp}[thm]{Proposition}
\newtheorem{lem}[thm]{Lemma}

\theoremstyle{definition}
\newtheorem*{rem}{Remark}

\begin{document}
\title{Finite-size scaling functions for directed polymers confined between attracting walls}
\author{A. L. Owczarek$^1$, T. Prellberg$^2$ and A. Rechnitzer$^3$\thanks{{\tt {\rm email:}
aleks@ms.unimelb.edu.au,t.prellberg@qmul.ac.uk,andrewr@math.ubc.ca}} \\
         $^1$Department of Mathematics and Statistics,\\
         The University of Melbourne,\\
         Parkville, Victoria 3052, Australia.\\
$^2$School of Mathematical Sciences\\ 
Queen Mary, University of London\\
Mile End Road, London E1 4NS, UK.\\
$^3$Department of Mathematics\\
University of British Columbia\\
Vancouver, BC, V6T-1Z2, Canada.\\
}
\date{\today}

\maketitle 

\begin{abstract}
  The exact solution of directed self-avoiding walks confined to a
  slit of finite width and interacting with the walls of the slit via
  an attractive potential has been calculated recently. The walks can
  be considered to model the polymer-induced steric stabilisation and
  sensitised floculation of colloidal dispersions.  The large width
  asymptotics led to a phase diagram different to that of a polymer
  attached to, and attracted to, a single wall. The question that
  arises is: can one interpolate between the single wall and two wall
  cases?
 
  In this paper we calculate the \emph{exact} scaling functions for
  the partition function by considering the two variable asymptotics
  of the partition function for simultaneous large length and large
  width. Consequently, we find the scaling functions for the force
  induced by the polymer on the walls. We find that these scaling functions
  are given by elliptic $\vartheta$-functions. In some parts of the
  phase diagram there is more a complex crossover between the single
  wall and two wall cases and we elucidate how this happens.
 
 \end{abstract}

\section{Introduction}
The problem of a single polymer confined between two walls and
interacting with those walls has been considered for at least 35 years
\cite{dimario1971a-a}. One reason for this is its use as a model of
the stabilization of colloidal dispersions by adsorbed polymers
(steric stabilization) and the destabilization when the polymer can
adsorb on surfaces of different colloidal particles (sensitized
flocculation). Until recently, even when one substitutes directed
walks for the more canonical self-avoiding walks in a two-dimensional
lattice model of this phenomenon, the only cases to be considered
exactly have been special cases where the interaction with the two
surfaces are equal. Recently, Brak \emph{et al.\ }\cite{brak2005a-:a} have
calculated the generating functions for a directed self-avoiding walk
confined by two horizontal walls on the square lattice where separate
Boltzmann weights $a$ and $b$ were associated with visits to the lower
wall and upper wall respectively, with various restrictions on the
end-points of the walks.

\begin{figure}[ht!]
\begin{center}
\includegraphics[height=7cm]{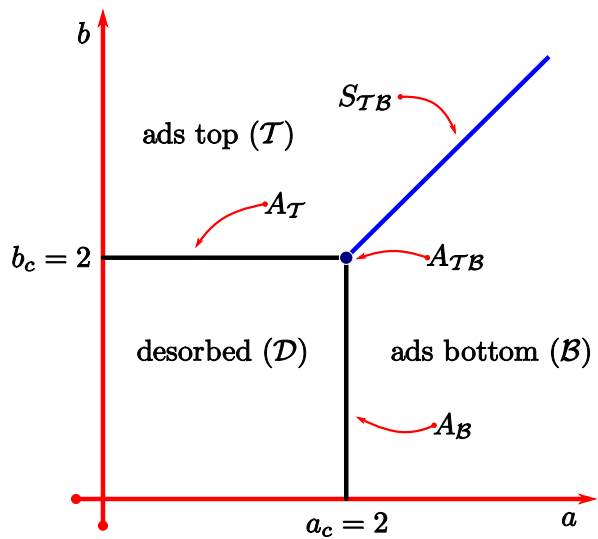}
\caption{\it  Phase diagram of the infinite strip where the separation
  of the walls is made large after the limit of infinite walk length
  is taken. There are 3 phases: desorbed (des), adsorbed onto the
  bottom wall (ads bottom) and adsorbed onto the top (ads top). Our
  notation for the various regions and transition lines are marked.
  The transition lines at $a=2,b\leq2$ and $b=2, a\leq2$ marking the
  boundary of the desorbed region are second order phase transitions
  with a jump in the specific heat on crossing the line while the line
  marking the boundary at $a=b,a>2$ of the two adsorbed regions is a
  first order transition.}
\label{phase-diagram} 
\end{center}
\end{figure}

\begin{figure}[ht!]
\begin{center}
\includegraphics[height=6cm]{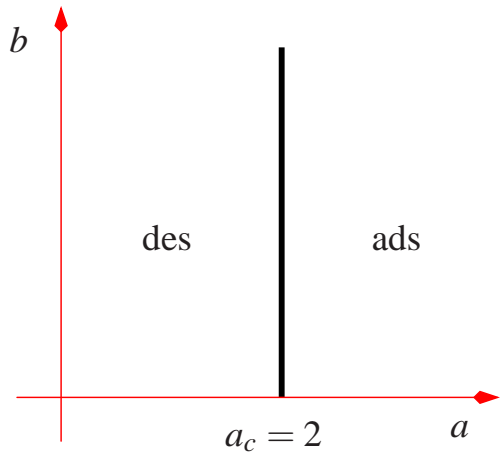}
\caption{\it  Phase diagram of the half-plane (one wall) problem 
  where the separation of the walls is made large \emph{before} the
  limit of infinite walk length is taken. There are 2 phases: desorbed
  (des) and adsorbed onto the bottom wall (ads bottom). The boundary
  of the two phase is a second order phase transition.}
\label{half-plane} 
\end{center}
\end{figure}

The dominant singularity of the generating function of any of the
subcases considered by Brak \emph{et al.\ }\cite{brak2005a-:a} leads to the
calculation of the free energy and the force induced by the polymer to the
walls. By considering the dominant singularity one is effectively considering
the infinite walk length limit. The dominant singularities of the
generating functions were analysed asymptotically for large widths.
In the infinite width limit a novel phase diagram was obtained, different to the
one obtained by analysing a polymer in a half-plane geometry.  In
Figure~\ref{phase-diagram} the phase diagram obtained by Brak \emph{et al.\
}\cite{brak2005a-:a} is given. For small $a$ and $b$ the polymer is
desorbed from both walls while for large $a$, respectively large $b$,
the polymer is absorbed onto one wall or the other. It is interesting
to note that new numerical results
\cite{rensburg2005a-:a,martin2007a-:a} and rigorous results
\cite{rensburg2006a-a} show that
undirected self-avoiding walks demonstrate very similar behaviour to
the exact solution of the directed walk model.
In the infinite width limit, referred to as the \emph{infinite slit}
(in two dimensions), the (reduced) free energy $\kappa(a,b)$ is given by
  \begin{equation}
    \kappa^{is}(a,b) =
    \begin{cases}
      \log(2) & a,b \leq 2 \\
      \log\left(\frac{a}{\sqrt{a-1}}\right) & a > 2 \mbox{ and } a>b \\
      \log\left(\frac{b}{\sqrt{b-1}}\right) &  b> 2 \mbox{ and } a<b
    \end{cases}
  \end{equation}
In the half-plane geometry the limit of infinite wall separation
is effectively taken before the limit of infinite walk length
(see Figure~\ref{half-plane}). The results in
\cite{brak2005a-:a} imply, unusually, that the order of the two
limits, polymer length to infinity and wall separation to
infinity, are not interchangable.  In fact, the free energy
depends only upon the value of $a$ and is given by
  \begin{equation}
    \kappa^{hp}(a,b) =
    \begin{cases}
      \log(2) & a \leq 2 \\
      \log\left(\frac{a}{\sqrt{a-1}}\right) & a > 2
    \end{cases}
  \end{equation}
Hence, for $b> 2 \mbox{ and } a<b$ (denoted region ${\cal T}$ in
Figure~\ref{phase-diagram}) the infinite slit and half-plane
free energies are different.  The question that naturally arises
is whether there is a scaling function that interpolates
between these two limits and whether this extends to the region
where the free energies differ. To do this one needs to
consider the finite length partition functions rather than the
generating functions.  An exact expression for the finite length
partition function has now been calculated \cite{brak2006a-:a}.
However, it is not easy to see how one could analyse this
expression asymptotically, especially for large separations.
Returning to the undirected model, in three dimensions, a
scaling theory \cite{martin2007a-:a} valid for the desorbed
region ${\cal D}$ of the infinite slit and its boundaries has
been shown (numerically) to hold. 

In this paper we calculate the two variable asymptotics for large wall
separation and large polymer length of the partition function of one
of the directed walk models considered by Brak \emph{et al.\ 
}\cite{brak2005a-:a}. The paper is set out as follows. In Section 2 we
set the stage for our calculations: we recall the exact definition of
the model and the calculated generating function, and introduce the
notation which we will use in the remainder of the paper, in
particular, a parametrisation in more convenient variables.  Much of
the analysis hinges on the understanding of the singularities of the
generating function; this will be discussed in Section 3.  In Section
4 we show how the contour integral expression for the partition
function can be reformulated in terms of residues of the singularities
of the generating function. We discuss special cases for specific
values of $a$ and $b$, where one obtains simple expressions for the
partition function, and also give an exact expression for general $a$
and $b$.  In Section 5 we calculate the scaling function of the
partition function for the special cases, and then derive 
the scaling function in the various regions of the phase diagram for
general $a$ and $b$. In Section 6 we discuss the results in the light
of finite size scaling theory and, in particular, discuss how in each
region of the phase diagram the scaling results interpolate between
the single and double wall models. Importantly, we demonstrate that in
the desorbed region of the infinite slit and on its boundaries the
scaling theory proposed for undirected SAW in a three-dimensional slab
holds exactly for our directed model, with appropropiate exponent
substitutions.

\section{The model}
\begin{figure}[ht!]
\begin{center}
\includegraphics[height=5cm]{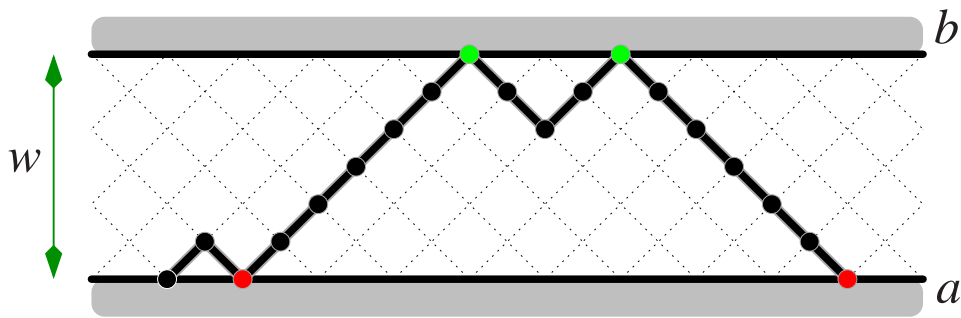}
\caption{\it  An example of a directed path which is a loop: both ends
  of the walk are fixed to be on the bottom wall. A Boltzmann weight
  $a$ is associated with visits to the bottom wall (excluding the first) and a
  Boltzmann weight $b$ is associated with visits to the
  top wall.}
\label{loop} 
\end{center}
\end{figure}

Brak \emph{et al.\ }\cite{brak2005a-:a} considered three different
end-point restrictions.  In this paper we shall restrict ourselves to
the case when both end-points are attached to the same wall: in
\cite{brak2005a-:a} these were referred to as \emph{loops} (see
Figure~\ref{loop}).
If $\mathcal{L}_w^n$  is the
set of loops of fixed length $n$ edges in the slit of
width $w$ then the partition function of loops is defined as
\begin{equation}
\label{part-def}
Z_{n,w}(a,b)  =   \sum_{p \in \mathcal{L}_w^n} a^{u(p)} b^{v(p)}
\end{equation}
where $u(p)$ and $v(p)$ are the number of vertices in the line $y=0$
(excluding the zeroth vertex) and the number of vertices in the line
$y=w$, respectively.  The generating function $L_w(z,a,b)$ is then
given by
\begin{equation}
\label{genfn-def}
  L_w(z,a,b)  =   \sum_{n=0}^\infty Z_{n,w}(a,b) \; z^{n} .
\end{equation}
The force ${\cal F}_{n,w}(a,b)$ is defined as 
\begin{equation}
{\cal F}_{n,w}(a,b)
  =  \frac{\log\left(Z_{n,w+1}(a,b)\right)
  - \log\left(Z_{n,w}(a,b)\right) }{n}
\end{equation}
but can be estimated from an asymptotic expression for large $w$ as
\begin{equation}
{\cal F}_{n,w} = \frac{1}{n \, Z_{n,w}}\frac{\partial Z_{n,w}}{\partial w}
\end{equation}

The aim of this paper is to extract the asymptotics of the finite size partition
function $Z_{n,w}(a,b)$ by inverting equation~(\ref{genfn-def}). Hence we have
\begin{equation}
\label{inverse}
Z_{n,w}(a,b)=\frac1{2\pi i}\oint L_w(z,a,b)\frac{dz}{z^{n+1}}\;,
\end{equation}
where the generating function $L_w(z,a,b)$ has been calculated in
\cite{brak2005a-:a} and is
\begin{equation}
L_w(z,a,b)=\frac{(1+q)[(1+q-bq)-(1+q-b)q^w]}{(1+q-aq)(1+q-bq)-(1+q-a)(1+q-b)q^w}
\;,
\end{equation}
with $z=\sqrt q/(1+q)$. The problem is to evaluate the above contour integral
for large but finite $n$ and $w$. 

Before we enter the calculations let us introduce some notation to
keep track of the different regions and transitions in the phase
diagram of the infinite slit (Figure~\ref{phase-diagram}). We restrict
our discussion to $a,b\geq1$ and label the \emph{desorbed} region, $a<
2, b < 2$ as ${\cal D}$, the region where the polymer \emph{adsorbs
  onto the bottom wall}, $a>2,a>b$ as ${\cal B}$, and the region where
the polymer \emph{adsorbs onto the top wall}, $b>2,b>a$ as ${\cal T}$.
The boundary where the region ${\cal D}$ meets region ${\cal B}$,
$a=2, b<2$, that is, when the polymer is critically adsorbing onto the
botttom wall, we denote as $\textrm{A}_{\cal B}$. The boundary where
the region ${\cal D}$ meets region ${\cal T}$, $b=2, a<2$, that is,
when the polymer is critically adsorbing onto the top wall, we denote
as $\textrm{A}_{\cal T}$.  The boundary where the region ${\cal B}$
meets region ${\cal T}$, $a=b, a>2$, that is, when the polymer is
equally adsorbed onto the both walls, we denote as $\textrm{S}_{\cal
  TB}$. Finally, the point where all three regions and all three lines
meet at $a=b=2$ is denoted $\textrm{A}_{\cal TB}$.

Mathematically it is advantageous to re-parametrise $L_w$ by introducing
$a=1+\lambda^2$, $b=1+\mu^2$ with $\lambda,\mu\geq0$ and $q=p^2$. Hence $z = p/(1+p^2)$. In
what follows, we will thus work with 
\beq\label{workwith}
L_w(z,a,b)=\Lc_w(p,\lambda,\mu)=\frac{(1+p^2)[(1-\mu^2p^2)+
(\mu^2-p^2)p^{2w}]}{(1-\lambda^2p^2)(1-\mu^2p^2)-(\lambda^2-p^2)(\mu^2-p^2)p^{2w}}\;,
\eeq 
so that equation~(\ref{inverse}) becomes
\beq
\label{cauchy}
Z_{n,w}(a,b)=\frac1{2\pi i}\oint \Lc_w(p,\lambda,\mu)(1-p^2)(1+p^2)^{n-1}\frac{dp}{p^{n+1}}\;.
\eeq

\section{Singularities}

Of crucial importance for the understanding of the structure of the generating function
are its singularities, i.e. the zeros of the denominator polynomial
\beq
\label{denominator}
    D_w(p) =(1-\lambda^2p^2)(1-\mu^2p^2)-(\lambda^2-p^2)(\mu^2-p^2)p^{2w}\;.
\eeq
It is convenient to look at some special cases first, where $D_w(p)$ simplifies considerably.
We have
\beq
\label{specialzeros}
D_w(p)=
\left\{\begin{array}{cl}
1-p^{4+2w}
&\mbox{for $\lambda=0$, $\mu=0$}\;,\\
(1-p^2)(1+p^{2+2w})
&\mbox{for $\lambda=0$, $\mu=1$ or $\lambda=1$, $\mu=0$}\;,\\
(1-\lambda^2p^2)(1-\lambda^{-2}p^2)(1-p^{2w})
&\mbox{for $\lambda\mu=1$}\;.
\end{array}
\right.
\eeq
For these special cases, all zeros are simple with the exception of the case $\lambda=1$ and $\mu=1$, in which 
case $p=\pm1$ is a multiple zero. For general values of $\lambda$ and $\mu$, 
we have the following result on the multiplicity of zeros.

\begin{lem}
\label{multiplicity}
  If $(\lambda,\mu)\neq(1,1)$, the polynomial
  \begin{equation}
    D_w(p) =(1-\lambda^2p^2)(1-\mu^2p^2)-(\lambda^2-p^2)(\mu^2-p^2)p^{2w}
  \end{equation}
  has simple zeros except possibly at $p=\pm 1$ for a single value of $w$.
\end{lem}
\proof The polynomial $D_w(p)$ is simply related to the orthogonal
polynomials, $P_w(z)$, defined in \cite{brak2005a-:a}. In particular
\begin{equation}
  D_w(p) =  (1-p^2) (1+p^2)^{w+1} P_w\left(\frac{p}{1+p^2}\right).
\end{equation}
Note that $P_w(z)$ is a polynomial of degree $w+1$ in $z$, so the
above expression is indeed polynomial in $p$. The zeros, $p_i$, of
$D_w(p)$ are therefore either $p = \pm 1$ or images of the zeros,
$z_i$, of $P_w(z)$ given by
\begin{equation}\label{trafo}
  p = \frac{1 \pm \sqrt{1-4z_i^2}}{2z_i}
\end{equation}
A standard result on orthogonal polynomials (see Theorem 5.4.1 in
\cite{andrews1999} for example) implies that the zeros of $P_w(z)$ are
simple. Hence the images of these zeros under the above mapping are
simple, except possibly when $z = \pm1/2$. If $P_w(z)$ has a zero at
$z=\pm1/2$, then it follows that $D_w(p)$ may have multiple zeros at
$p=\pm1$.

We now show that such multiple zeros at $p=\pm1$ can only occur for a
single value of $w$. The derivative of $D_w(p)$ at $p=\pm1$ is given by
\begin{equation}
  D_w'(\pm1) = \mp 4 (\lambda^2 \mu^2 -1) \pm 2 (\lambda^2 -1) (\mu^2- 1) w.
\end{equation}
If $(\lambda,\mu)\neq(1,1)$ then this derivative is
zero when
\begin{equation}
  w = 2\frac{\lambda^2 \mu^2 - 1}{(\lambda^2-1)(\mu^2-1)}.
\end{equation}
Hence it is only at this single value of $w$ that $D_w(p)$ can have
multiple zeros at $p=\pm1$.

\qed

Under the transformation~(\ref{trafo}), the zeros of $D_w(p)$ are related to the
zeros of the orthogonal polynomial $P_w(z)$, which are all real. The
transformation from $z$ to $p$ then implies that $D_w(p)$ can only have roots on
the unit circle or the real axis.  Thus it makes sense to introduce the
parametrisation $p=e^{it}$ for roots on the unit circle. A straightforward
calculation leads to the results summarised in the next lemma.

\begin{lem}
The zeros of the polynomial $D_w(p)$ 
are given by $p_k=e^{it_k}$, where
\beq\label{tant}
\tan wt_k=
\frac
{\displaystyle \left(\frac{\lambda^2-1}{\lambda^2+1}
      +\frac{\mu^2-1}{\mu^2+1}\right)\tan t_k}
{\left(\displaystyle \frac{\lambda^2-1}{\lambda^2+1}\right)
 \left(\displaystyle \frac{\mu^2-1}{\mu^2+1}\right)-\tan^2t_k } \;.
\eeq
Additionally, for $\lambda=1$ or $\mu=1$, $D_w(p)=0$ at $p=\pm1$. If $\lambda=\mu=1$, $p=\pm1$ is a triple zero of $D_w(p)$.
In particular, for $w$ sufficiently large, the numbers of zeros of $D_w(p)$, including their multiplicity, are given as follows:
\begin{enumerate}
\item
If $\lambda<1$ and $\mu<1$, equation~(\ref{tant}) has $2w+4$ real solutions.
$D_w(p)$ has $2w+4$ zeros on the unit circle.
\item
On the line segments $\lambda<1$ and $\mu=1$, respectively $\lambda=1$ and
$\mu<1$, equation~(\ref{tant}) has $2w+2$ solutions. $D_w(p)$ has $2w+4$ zeros
on the unit circle.
\item
If $\lambda=\mu=1$, equation~(\ref{tant}) has $2w$ solutions.
$D_w(p)$ has $2w+4$ zeros on the unit circle.
\item
If $\lambda>1$ and $\mu<1$ or $\lambda<1$ and $\mu>1$, equation~(\ref{tant}) has
$2w$ solutions. $D_w(p)$ has $2w$ zeros on the unit circle and $4$ zeros on the
real line.
\item
On the line segments $\lambda>1$ and $\mu=1$, respectively $\lambda=1$ and
$\mu>1$, equation~(\ref{tant}) has $2w-2$ solutions. $D_w(p)$ has $2w$ zeros on 
the unit circle and $4$ zeros on the real line.
\item
If $\lambda>1$ and $\mu>1$, equation~(\ref{tant}) has $2w-4$ solutions.
$D_w(p)$ has $2w-4$ zeros on  the unit circle and $8$ zeros on the real line.
\end{enumerate}
This corresponds to $(2w+4)-4\sigma$ zeros $p_k$ of $D_w(p)$ on the unit circle, where $\sigma\in\{0,1,2\}$, depending on the
values of $\lambda$ and $\mu$. The other $4\sigma$ zeros $p_k$ are located on the real line. 
If $p_k$ is a real zero, then so is $-p_k$, $1/p_k$, and $-1/p_k$.
\end{lem}

\proof

Equation~(\ref{tant}) is obtained from $D_w(p)$ by substitution of $p=e^{it}$,
followed by routine simplification. The number of real solutions is most easily
obtained by considering the graphs of the LHS and RHS of~(\ref{tant}) over the
interval $[0,2\pi)$. The function $\tan wt_k$ is monotonic and has $2w$
simple poles, whereas the behaviour of the RHS depends on the values of
$\lambda$ and $\mu$. For example, if both $\lambda<1$ and $\mu<1$, the RHS is
monotonically decreasing and has $4$ poles, leading to a total number of $2w+4$
intersections of both graphs. The other cases can be obtained similarly.

\endproof

\begin{rem}
The real roots $p_k$ can be obtained correspondingly from $p_k=e^{s_k}$, where now
\beq
\tanh ws_k=
\frac
%
{\left({\displaystyle \frac{\lambda^2-1}{\lambda^2+1}
      +\frac{\mu^2-1}{\mu^2+1}} \right)\tanh s_k}
{\displaystyle
\left(\frac{\lambda^2-1}{\lambda^2+1}\right)
 \left(\frac{\mu^2-1}{\mu^2+1}\right)+\tanh^2s_k}\;.
\eeq

\end{rem}

\section{Partition function identities}

In this section we derive explicit expressions for the partition function which are especially suited
for an asymptotic analysis of the finite-size scaling behaviour. The key is the following Lemma.
\begin{lem}
\label{key}
\beq\label{residues}
Z_{n,w}(a,b)=-\frac12\sum_{p_k}\mathrm{Res}\left(f;p_k\right)\;,
\eeq
where
\beq
\label{label_f}
f(p)=\frac{(1-p^2)[(1-\mu^2p^2)+(\mu^2-p^2)p^{2w}]}{D_w(p)}\left(\frac{1+p^2}p\right)^n\frac1p
\eeq
and $p_k$ are the zeros of $D_w(p)$.
\end{lem}

\proof
From equation~(\ref{cauchy}) it follows that
\beq\label{thisisit}
Z_{n,w}(a,b)=\frac1{2\pi i}\oint f(p)dp\;,
\eeq 
with $f(p)$ given in equation~(\ref{label_f}).
The integrand is a rational function in $p$ so that the sum over all its residues on the Riemann
sphere is equal to zero. From this it follows that 
\beq
Z_{n,w}(a,b)=\mathrm{Res}\left(f;0\right)=-\sum_{p_k}\mathrm{Res}\left(f;p_k\right)-\mathrm{Res}\left(f;\infty\right)\;,
\eeq where the $p_k$ are the zeros of $D_w(p)$.


Due to the symmetry
\beq
f(1/p)=-p^2f(p)
\eeq
the singularities $p_k$ come in pairs $(p_k,1/p_k)$ and the paired residues are equal, i.e.
\beq
\mathrm{Res}\left(f;p_k\right)=\mathrm{Res}\left(f;1/p_k\right)\;.
\eeq
It therefore follows that
\beq
Z_{n,w}(a,b)=\mathrm{Res}\left(f;0\right)=-\frac12\sum_{p_k}\mathrm{Res}\left(f;p_k\right)\;.
\eeq

\qed

It is convenient to first look at the special cases already considered above
(see equation~(\ref{specialzeros})).
\begin{prp}
We have
\begin{subequations}
\begin{align}
Z_{n,w}(1,1) & =
\frac{2^{n+1}}{w+2}\sum_{k=1}^{w+1}\sin^2\frac{k\pi}{w+2}\cos^n\frac{k\pi}{w+2}
\;,\\  
Z_{n,w}(1,2) &= 
\frac{2^n}{w+1}\sum_{k=0}^{2w+1}\sin^2\frac{(2k+1)\pi}{2w+2}\cos^n\frac{(2k+1)\pi}{2w+2}
\;,\\
Z_{n,w}(2,1) & = 
\frac{2^n}{2(w+1)}\sum_{k=0}^{2w+1}\cos^n\frac{(2k+1)\pi}{2w+2}
\;,\\
Z_{n,w}(2,2) & = \frac{2^n}{w}\sum_{k=0}^{w-1}\cos^n\frac{k\pi}w
\;,
\end{align}
and
\begin{equation}
  \begin{split}
    Z_{n,w}\left(1+\lambda^2,1+\lambda^{-2}\right) =
    \frac{1-\lambda^2}{2\lambda^2} \frac{\lambda^{2w}}{1-\lambda^{2w}}
    \left(\lambda+\lambda^{-1}\right)^n \left(1+(-1)^n\right)\\
    +\frac{2^n}{\lambda w} \sum_{k=1}^{w-1}\frac
    {\frac{\lambda+\lambda^{-1}}2\sin^2\frac{k\pi}w}
    {(\frac{\lambda-\lambda^{-1}}2)^2+\sin^2\frac{k\pi}w}
    \cos^n\frac{k\pi}w
  \end{split}
\end{equation}
\end{subequations}
for $\lambda\neq1$.
\end{prp}

\proof
Equation~(\ref{specialzeros}) allows for an explicit calculation of the zeros of
$D_w(p)$. An application of Lemma~\ref{key} gives the result.

\qed

We note that the expression for $Z_{n,w}(1,1)$ is a special case of results
in \cite{krattenthaler2003}. For general values of $\lambda$ and $\mu$, we
arrive at the following result.
\begin{prp}
For $\lambda\mu\neq1$ and $w$ sufficiently large,
\beq\label{identity}
Z_{n,w}(a,b)=\frac{1+\lambda^2}4\sum_{p_k}
\frac{(1-p_k^2)^2}{(\lambda^2-p_k^2)(1-\lambda^2p_k^2)}\left(\frac{1+p_k^2}{p_k}\right)^n\frac1{w+\varepsilon_k}\;,
\eeq
where $p_k$ are the zeros of $D_w(p)$
and
\beq\label{epsilon}
\varepsilon_k=p_k^2\left[\frac{\lambda^2}{1-\lambda^2p_k^2}-\frac1{\lambda^2-p_k^2}+\frac{\mu^2}{1-\mu^2p_k^2}-\frac1{\mu^2-p_k^2}\right]\;.
\eeq
\end{prp}

\proof
Using Lemma~\ref{key}, we write
$f(p)=r(p)/D_w(p)$ with
\beq
\label{rofp}
r(p)=(1-p^2)[(1-\mu^2p^2)+(\mu^2-p^2)p^{2w}]\left(\frac{1+p^2}p\right)^n\frac1p\;.
\eeq
According to Lemma~\ref{multiplicity}, all roots of $D_w(p)$ are simple for $w$
sufficiently large, and $p=\pm1$ are removable singularities of $f(p)$ due to
the occurrence of the factor $(1-p^2)$ in $r(p)$.
Using that
\beq
\mathrm{Res}\left(f;p_k\right)=r(p_k)/D_w'(p_k)
\eeq
for simple poles at $p_k$, we find that
\beq
  \label{residue}
  \mathrm{Res}\left(f;p_k\right) = 
  -\frac{(1-p_k^2)[(1-\mu^2p_k^2)+(\mu^2-p_k^2)p_k^{2w}]\left(\frac{1+p_k^2}{p_k}\right)^n}
  {2w(\lambda^2-p_k^2)(\mu^2-p_k^2)p_k^{2w}+2p_k^2[(\lambda^2+\mu^2-2\lambda^2\mu^2p_k^2)-(\lambda^2+\mu^2-2p_k^2)p_k^{2w}]}\;.
\eeq
Eliminating $p_k^{2w}$ using $D_w(p_k)=0$ gives
\beq
\mathrm{Res}\left(f;p_k\right)=-\frac{1+\lambda^2}2\frac{(1-p_k^2)^2}{(\lambda^2-p_k^2)(1-\lambda^2p_k^2)}\left(\frac{1+p_k^2}{p_k}\right)^n\frac1{w+\varepsilon_k}\;.
\eeq
where $\varepsilon_k$ is given by~(\ref{epsilon}).
\endproof

\begin{rem}
The condition $\lambda\mu\neq1$ is necessary because when $\lambda\mu=1$ a root,
$p_k$, is located at $p_k=\lambda$, making the expression for $Z_{n,w}(a,b)$
invalid as it is written. However, in this case we already have given a much
more explicit expression for $Z_{n,w}(a,b)$ above.
\end{rem}

\section{Asymptotics}

If we consider the scaling behaviour of long walks in wide strips, the behaviour
of our model should depend on the relation between average vertical displacement
of an unrestricted polymer and the width of the strip, $w$. We therefore
expect the occurrence of the scaling combination $\sqrt n/w$ in the asymptotics,
and this is indeed what a mathematical analysis shows.

As in the previous sections, it is convenient to first consider the special
cases.
\begin{prp}
For $n$ even, $\sqrt n/w$ fixed and $n\rightarrow\infty$,
\begin{subequations} 
\begin{align}
Z_{n,w}(1,1) & \sim\frac{2^n}{n^{3/2}}f_{0,0}(\sqrt n/w) \;, \\
Z_{n,w}(1,2) & \sim\frac{2^n}{n^{3/2}}f_{0,1}(\sqrt n/w) \;, \\
Z_{n,w}(2,1) & \sim\frac{2^n}{n^{1/2}}f_{1,0}(\sqrt n/w) \;, \\
Z_{n,w}(2,2) & \sim\frac{2^n}{n^{1/2}}f_{1,1}(\sqrt n/w) \;, \\
\intertext{for $\lambda<1$ we have}
Z_{n,w}\left(1+\lambda^2,1+\lambda^{-2}\right) & \sim 
\frac{1-\lambda^2}{\lambda^2}\lambda^{2w}\left(\lambda+\lambda^{-1}\right)^n
+\frac{2^n}{n^{3/2}}f_{\lambda,\lambda^{-1}}(\sqrt n/w)\;, \\
\intertext{while for $\lambda>1$ we have}
Z_{n,w}\left(1+\lambda^2,1+\lambda^{-2}\right) & \sim
\frac{\lambda^2-1}{\lambda^2}\left(\lambda+\lambda^{-1}\right)^n
+\frac{2^n}{n^{3/2}}f_{\lambda,\lambda^{-1}}(\sqrt n/w).
\end{align}
\end{subequations}
The functions $f_{\lambda,\mu}$ are given by
\begin{subequations}
\begin{align}
f_{0,0}(x) &= 
2\pi^2x^3\sum_{k=-\infty}^\infty k^2e^{-\frac{\pi^2k^2}2x^2}
=2\pi^2x^3e^{-\frac{\pi^2x^2}2} \;
\vartheta_3'\left(e^{-\frac{\pi^2x^2}2}\right) \;,\\
f_{0,1}(x) &=
2\pi^2x^3\sum_{k=-\infty}^\infty (k+1/2)^2e^{-\frac{\pi^2(k+1/2)^2}2x^2}
=2\pi^2x^3e^{-\frac{\pi^2x^2}2} \;
\vartheta_2'\left(e^{-\frac{\pi^2x^2}2}\right) \;, \\
f_{1,0}(x) &=
x\sum_{k=-\infty}^\infty e^{-\frac{\pi^2(k+1/2)^2}2x^2}=x
\; \vartheta_2\left(e^{-\frac{\pi^2x^2}2}\right)\;, \\
f_{1,1}(x) &=
x\sum_{k=-\infty}^\infty e^{-\frac{\pi^2k^2}2x^2}=x
\; \vartheta_3\left(e^{-\frac{\pi^2x^2}2}\right)\;,\\
\intertext {for $\lambda\neq1$,}
f_{\lambda,\lambda^{-1}}(x) &=
\frac{(\lambda^2+1)}{(\lambda^2-1)^2}2\pi^2x^3\sum_{k=0}^\infty k^2e^{-\frac{\pi^2k^2}2x^2}
=\frac{(\lambda^2+1)}{(\lambda^2-1)^2}2\pi^2x^3e^{-\frac{\pi^2x^2}2}
\; \vartheta_3'\left(e^{-\frac{\pi^2x^2}2}\right)
\;,
\end{align}
\end{subequations}
Here, $\vartheta_2(q)=\sum_{n=-\infty}^\infty q^{(n+1/2)^2}$ and
$\vartheta_3(q)=\sum_{n=-\infty}^\infty q^{n^2}$ are elliptic $\vartheta$-functions.

\end{prp}

\proof

We shall only discuss the case $\lambda=\mu=1$, as the argument applies \textit{mutatis mutandum} to the other cases.
Upon introducing the variable $x=n^{1/2}/w$, we can write
\beq
Z_{n,w}(2,2) = \frac{2^nx}{n^{1/2}}\sum_{k=0}^{w-1}
\cos^n\left(\frac{k\pi x}{n^{1/2}} \right)\;.
\eeq
For fixed $k$ we find that
\beq
\lim_{n\rightarrow\infty}
\cos^n\left(\frac{k\pi x}{n^{1/2}}\right)
=e^{-\frac{\pi^2k^2}2x^2}\;,
\eeq
and a similar contribution comes from the upper boundary of the summation for $k'=w-k$ fixed. The contribution to
the sum from other terms can be easily shown to be negligible, so that
\beq
\lim_{n\rightarrow\infty}\frac{n^{1/2}}{2^n}Z_{n,(n^{1/2}/x)}(2,2) =
x\sum_{k=-\infty}^\infty e^{-\frac{\pi^2k^2}2x^2}\;.
\eeq

\qed

\begin{figure}[ht!]
  \centering
  \begin{tabular}{cc}
  \includegraphics[width=7.3cm]{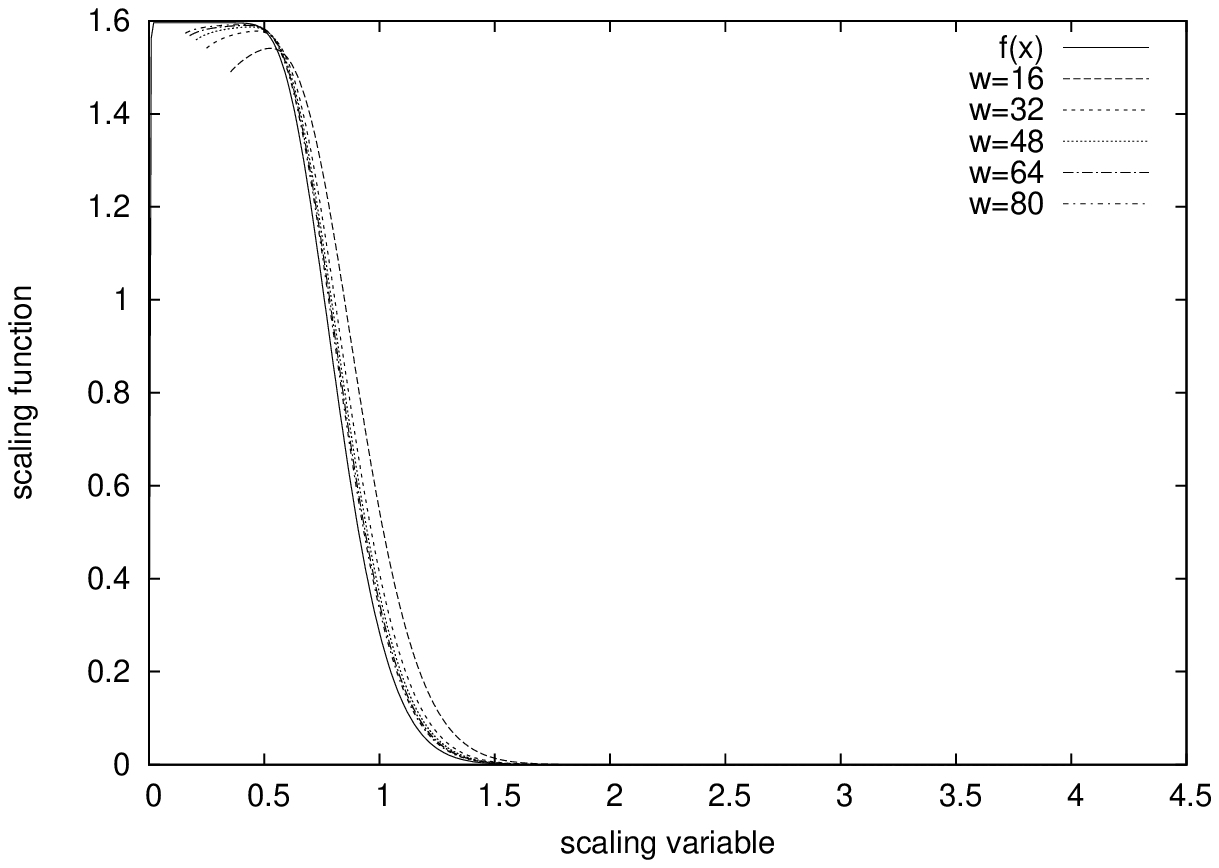}&
  \includegraphics[width=7.3cm]{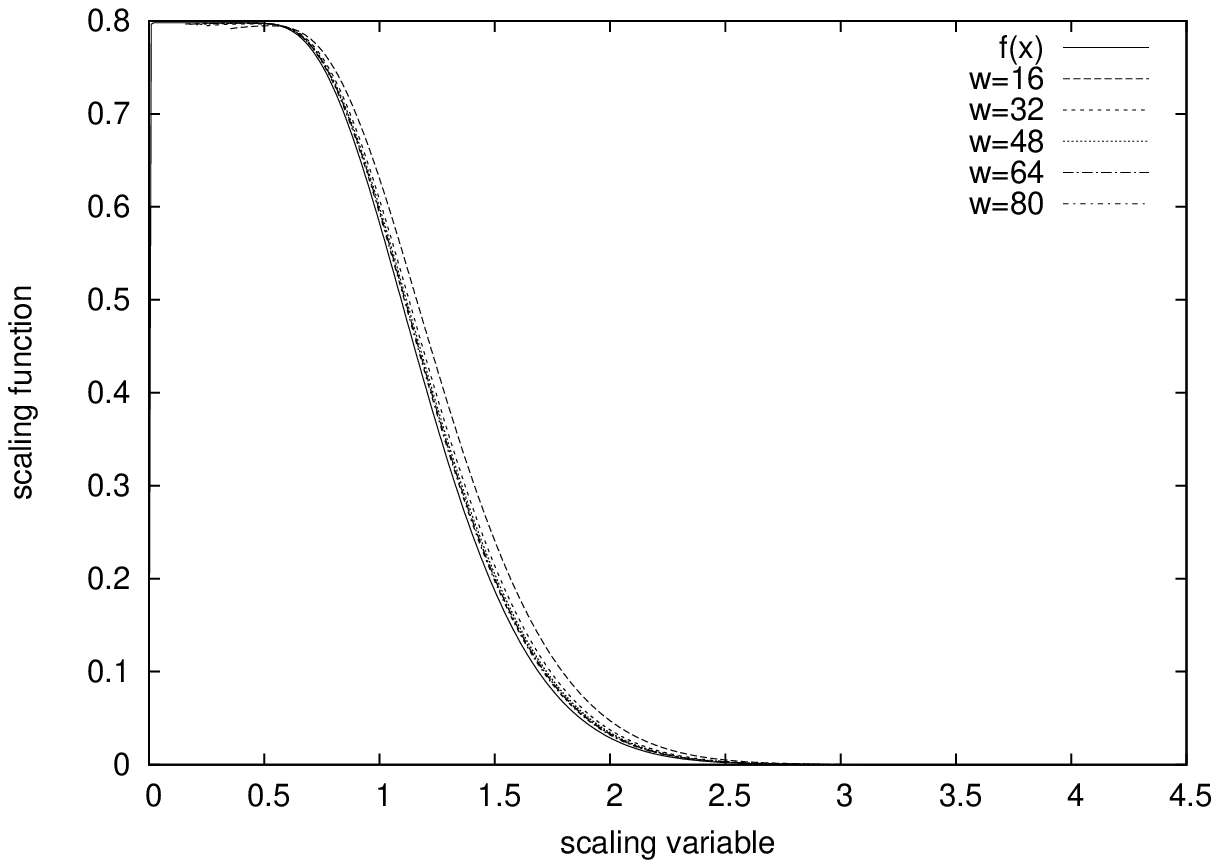}\\
  $(\mu,\lambda)=(0,0)$ & $(\mu,\lambda)=(1,0)$\\[3ex]
  \includegraphics[width=7.3cm]{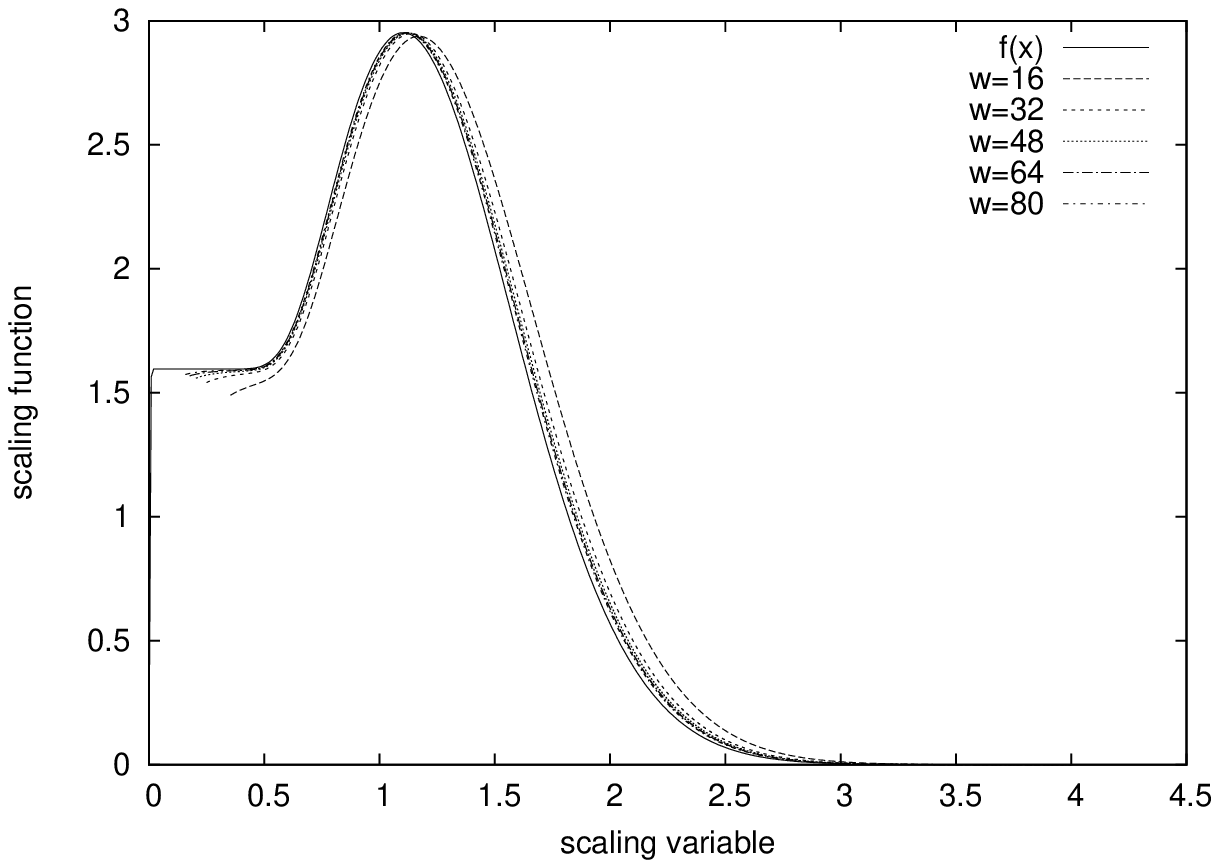}&
  \includegraphics[width=7.3cm]{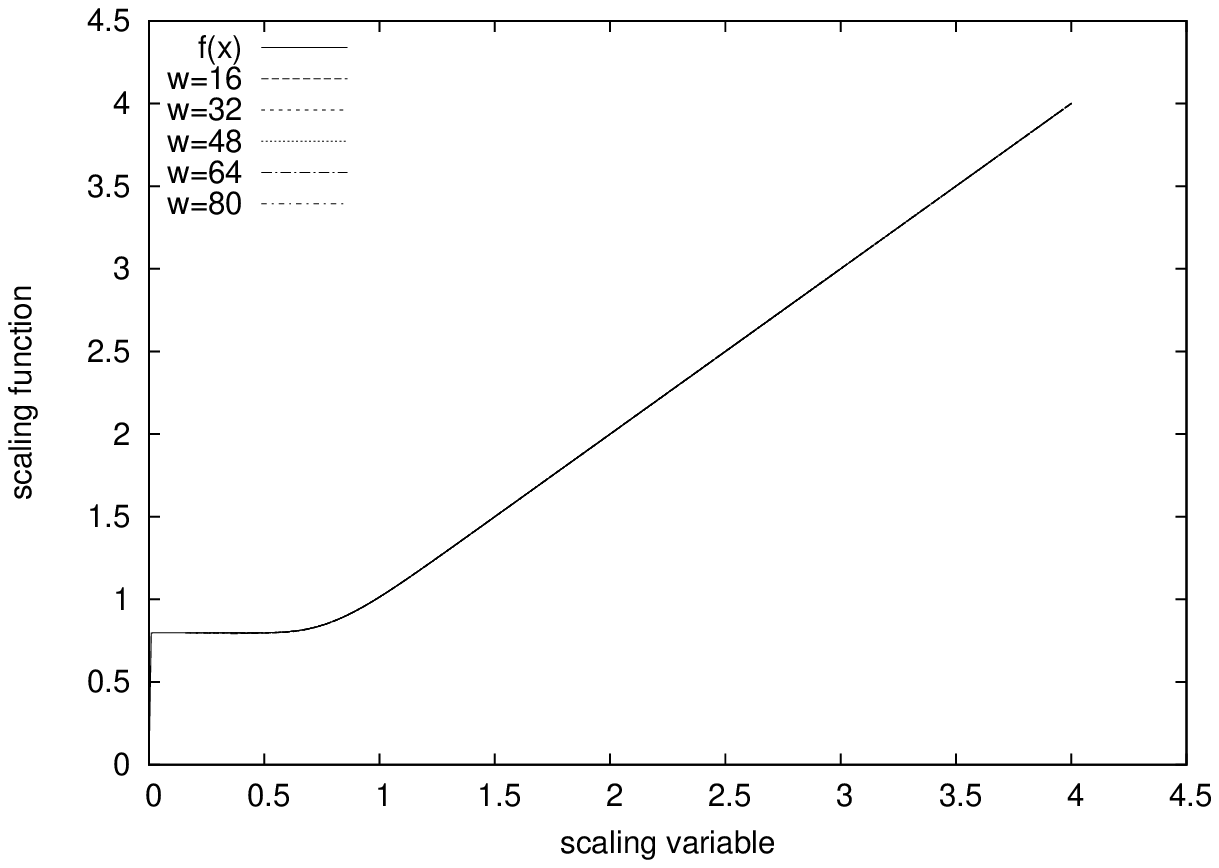}\\
  $(\mu,\lambda)=(0,1)$ & $(\mu,\lambda)=(1,1)$
  \end{tabular}
  \caption{\it The scaling functions (solid lines) at $(\mu,\lambda) = (0,0),
(1,0), (0,1)$ and $(1,1)$ respectively. Also plotted is numerical data
(dashed lines) for widths ranging from $16$ to $80$; these show that the
partition functions converge rapidly to the scaling forms.}
  \label{fig collapse}
\end{figure}

In Figure~\ref{fig collapse}  we have plotted the scaling functions at the
points $\lambda, \mu \in \{0,1\}$. These show that the partition functions
converge rapidly to the scaling functions even at quite modest widths ($w$
ranges from $16$ to $80$ and the maximum length is $16 w^2$). The
non-monotonic behaviour of $f_{01}(x)$ has also been observed in the three
dimensional self-avoiding walk model at a comparable point in the phase
diagram~\cite{martin2007a-:a}.

\begin{rem}
We note that even though we have only considered the scaling limit of $x=\sqrt
n/w$ fixed, the above scaling forms permit asymptotic matching to the limits
$w\rightarrow\infty$ with $n$ fixed, and $n\rightarrow\infty$ with $w$ fixed.
For example, we obtain
\beq
\frac{2^n}{n^{1/2}}f_{1,1}(\sqrt n/w)\sim\left\{
\begin{array}{ccc}
\frac1w2^n&\quad&n\rightarrow\infty\\[1ex]
\sqrt{\frac2{\pi n}}2^n&\quad&w\rightarrow\infty
\end{array}
\right.
\eeq
which agrees with the known asymptotic behaviour of $Z_{n,w}(2,2)$ for fixed $w$ or $n$, respectively.
\end{rem}

In order to discuss the case of general $a$ and $b$, let us consider
equation~(\ref{identity}) more closely. Note that for large $w$ the terms
$\varepsilon_k$ given by (\ref{epsilon}) become negligible in comparison with
$w$ if $p_k$ is a root on the unit circle, as $\varepsilon_k$ is then uniformly
bounded in $k$, $w$ and $n$. Also note that in this case the dependence
of~(\ref{identity}) on $\mu$ in enters only through the roots $p_k$.

However, if $p_k$ is a real root it needs to be treated individually, and 
we therefore split the partition function into two sums,
\begin{align}
  Z_{n,w}(a,b) & =Z_{n,w}^{(e)}(a,b)+Z_{n,w}^{(s)}(a,b)\;,\\
  \intertext{where}
  Z_{n,w}^{(e)}(a,b) & = 
  \frac{1+\lambda^2}{4} \sum_{|p_k|\neq1}
  \frac{(1-p_k^2)^2}{(\lambda^2-p_k^2)(1-\lambda^2p_k^2)}
  \left(\frac{1+p_k^2}{p_k}\right)^n\frac1{w+\varepsilon_k} \\
  \intertext{and}
  Z_{n,w}^{(s)}(a,b) &=
  \frac{1+\lambda^2}4\sum_{|p_k|=1}
  \frac{(1-p_k^2)^2}{(\lambda^2-p_k^2)(1-\lambda^2p_k^2)}
  \left(\frac{1+p_k^2}{p_k}\right)^n\frac1{w+\varepsilon_k}
\end{align}
Here, $Z_{n,w}^{(e)}(a,b)$ is a sum over zero, four, or eight terms, depending
on the value of $w$, $\lambda$, and $\mu$. While it turns out that
$Z_{n,w}^{(s)}(a,b)$ admits a scaling form, $Z_{n,w}^{(e)}(a,b)$ does not.
The following propositions give asymptotic estimates for $Z_{n,w}^{(e)}(a,b)$
and $Z_{n,w}^{(s)}(a,b)$ for large $w$.

The analysis of $Z_{n,w}^{(e)}(a,b)$ uses an asymptotic estimate of the zeros for which $p_k>1$, analogous to
equations~(7.7) and~(7.10) in \cite{brak2005a-:a}. If $\lambda>1$ and
$\mu\neq\lambda$, we find
\beq
p^2\sim\lambda^2+(\lambda^4-1)\left(
\frac{\lambda^2\mu^2-1}{\lambda^2-\mu^2} \right)
\lambda^ { -2w}\;,
\eeq
and if $\mu>1$ and $\lambda\neq\mu$, we find
\beq
p^2\sim\mu^2+(\mu^4-1)\left(\frac{\lambda^2\mu^2-1}{\mu^2-\lambda^2}\right)
\mu^{-2w } \; ,
\eeq
while for $\lambda=\mu>1$ we find
\beq
p^2\sim\lambda^2\pm(\lambda^4-1)\lambda^{-w}\;.
\eeq


\begin{figure}[ht!]
\begin{center}
\includegraphics[height=7cm]{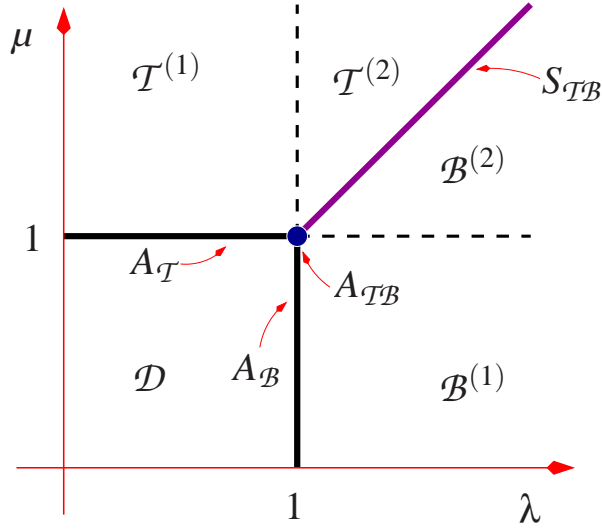}
\caption{\it  Regions that have different scaling forms. The definitions via
inequalities occur in equation~(\ref{regions}). The region ${\cal T}^{(1)}$
includes the dashed line segment $\lambda=1,\mu>1$ and the region ${\cal
B}^{(1)}$ includes the dashed line segment $\mu=1,\lambda >1$}
\label{finer} 
\end{center}
\end{figure}
In Figure~\ref{finer} we introduce a finer division of the regions of the phase
plane which will be needed below; essentially we split the regions ${\cal
T}$ and ${\cal B}$ into two parts. 

From the considerations above we arrive at the following result.

\begin{prp}
\label{prop-part-e-scale}
For $n$ even and $n\rightarrow\infty$,
\begin{multline}
\label{part-e-scale}
Z_{n,w}^{(e)}(a,b) 
 \sim
 \begin{cases}
   \ds
   \frac{(\lambda^2+1)(\mu^2-1)^2(\mu^4-1)}{\mu^2(\mu^2-\lambda^2)^2}
   \mu^{-2w}\left(\mu+\mu^{-1}\right)^n
   &\quad\text{in the region ${\cal T}^{(1)}$,}\\[3ex]
   \ds
   \left[
     \frac{(\lambda^2+1)(\mu^2-1)^2(\mu^4-1)}
     {\mu^2(\mu^2-\lambda^2)^2}\mu^{-2w}\left(\mu+\mu^{-1}\right)^n
     + \frac{\lambda^2-1}{\lambda^2}\left(\lambda+\lambda^{-1}\right)^n
   \right]
   &\quad\text{in the region ${\cal T}^{(2)}$,}\\[3ex]
   \ds
   \frac{\lambda^2-1}{\lambda^2}\left(\lambda+\lambda^{-1}\right)^n
   &\quad\text{on the line $\textrm{S}_{\cal TB}$,}\\[3ex]
   \ds
   \left[
     \frac{\lambda^2-1}{\lambda^2}\left(\lambda+\lambda^{-1}\right)^n  +
     \frac{(\lambda^2+1)(\mu^2-1)^2(\mu^4-1)}{\mu^2(\mu^2-\lambda^2)^2}
     \mu^{-2w}\left(\mu+\mu^{-1}\right)^n
   \right]
   &\quad\text{in the region ${\cal B}^{(2)}$, and}\\[3ex]
   \ds
   \frac{\lambda^2-1}{\lambda^2}\left(\lambda+\lambda^{-1}\right)^n
   &\quad\text{in the region ${\cal B}^{(1)}$.}
 \end{cases}
\end{multline}
where
\begin{equation}
  \begin{array}{lcl}
    \mathrm{S}_{\mathcal{TB}} = \{(\mu,\lambda)  \;|\;  \mu = \lambda  > 1\}\\
    \mathcal{T}^{(1)} = \{(\mu,\lambda) \;|\; \mu>1 \geq \lambda \} &
    \qquad &
    \mathcal{T}^{(2)} = \{(\mu,\lambda) \;|\;  \mu > \lambda > 1\} \\
    \mathcal{B}^{(1)} = \{(\mu,\lambda) \;|\; \lambda>1 \geq \mu \} &
    \qquad &
    \mathcal{B}^{(2)} = \{(\mu,\lambda) \;|\;  \lambda > \mu > 1\}.
  \end{array}
\label{regions}
\end{equation}
\end{prp}

The asymptotic estimate for $Z_{n,w}^{(s)}(a,b)$ is done in two stages. First, we present an estimate for large $w$ which is uniform
in $n$.

\begin{prp}
\beq
Z_{n,w}^{(s)}(a,b)=\left(\frac{\lambda+\lambda^{-1}}{4\lambda}\right)
\frac{2^n}{w}
\; \sum_{ t_k } \left(
\frac{\sin^2t_k}{\left(\frac{\lambda-\lambda^{-1}}2\right)^2+\sin^2t_k}
\cos^nt_k \right)\left[1+O(w^{-1})\right]
\eeq
uniformly in $n$, where $t_k$ are the roots of~(\ref{tant}) in $[0,\pi)$.
\end{prp}

\proof 
Apart from substituting $p_k=e^{it_k}$ and simplifying, the main effort lies in
obtaining an estimate for $\varepsilon_k$. For $\lambda,\mu\neq1$, we can
estimate it directly from~(\ref{epsilon})
\beq
|\varepsilon_k|<\frac{\lambda^2+1}{|\lambda^2-1|}+\frac{\mu^2+1}{|\mu^2-1|}\;.
\eeq
If $\lambda=1$, we find that
\beq
|\varepsilon_k|<\frac{\mu^2+1}{|\mu^2-1|}\;,
\eeq
and analogously for $\mu=1$. Finally, if both $\lambda=1$ and $\mu=1$ then $\varepsilon_k=0$.
\endproof

For the scaling behaviour of $Z_{n,w}^{(s)}(a,b)$, we find identical scaling
behaviour as in the special cases discussed above (except for
$\lambda$-dependent pre-factors).
\begin{prp}
For $n$ even, $\sqrt n/w$ fixed and $n\rightarrow\infty$,
\beq
\label{part-s-scale}
Z_{n,w}^{(s)}(a,b)\sim
\begin{cases}
\ds
\frac{2^n}{n^{3/2}}f(\sqrt n/w)
&\quad\text{for $\lambda\neq1$,}\\[2ex]
\ds
\frac{2^n}{n^{1/2}}f(\sqrt n/w)
&\quad\text{for $\lambda=1$.}\\[2ex]
\end{cases}
\eeq
where
\beq
\label{scale-fns}
f(x)=\begin{cases}
\ds
x \; \vartheta_3\left(e^{-\frac{\pi^2 x^2}2}\right)
&\quad\text{for $\lambda=1$ and $\mu=1$.}\\[2ex]
\ds
\frac{\lambda^2+1}{(\lambda^2-1)^2}
2\pi^2x^3e^{-\frac{\pi^2x^2}2}\; \vartheta_2'\left(e^{-\frac{\pi^2x^2}2}\right)
&\quad\text{for $\lambda\neq1$ and $\mu=1$,}\\[2ex]
\ds
x\; \vartheta_2\left(e^{-\frac{\pi^2 x^2}2}\right)
&\quad\text{for $\lambda=1$ and $\mu\neq1$, and}\\[2ex]
\ds
\frac{\lambda^2+1}{(\lambda^2-1)^2}
2\pi^2x^3e^{-\frac{\pi^2x^2}2}\; \vartheta_3'\left(e^{-\frac{\pi^2x^2}2}\right)
&\quad\text{for $\lambda\neq1$ and $\mu\neq1$.}
\end{cases}
\eeq
Here, $\vartheta_2(q)=\sum_{n=-\infty}^\infty q^{(n+1/2)^2}$ and
$\vartheta_3(q)=\sum_{n=-\infty}^\infty q^{n^2}$ are elliptic $\vartheta$-functions.

%
\end{prp}

\proof
The derivation is done in complete analogy to the special cases discussed above.  The only additional
consideration is that now $t_k$ is only known asymptotically for $w$ large.
If either $\lambda\neq1$ and $\mu\neq1$ or $\lambda=\mu=1$, then for large $w$
we find 
\beq
t_k\sim k\frac\pi w
\eeq
and otherwise (i.e. if $\lambda=1$ or $\mu=1$ but not $\lambda=\mu=1$) for
large $w$ we find 
\beq
t_k\sim(k+1/2)\frac\pi w\;.
\eeq
\endproof

We point out that these functions, which are effectively multiples of
the scaling functions (see below), are elliptic $\vartheta$-functions.
Further they are independent of $\lambda$ and $\mu$, apart from that overall
multiplicative factor, which is dependent only of $\lambda$ (as it depends on
the half-plane limit).  Finally we note that the result
for $\lambda=\mu=0$ is equivalent to asymptotic results of Flajolet
\emph{et al.\ }\cite{flajolet1993a-a} concerning the distribution of heights
of binary trees since there is a correspondence between directed paths
and binary trees.

\section{Discussion}
We now return to the issue of how the scaling forms calculated above
interpolate between the half-plane  and the infinite-slit. For the
half-plane we have (recalling $a=1+\lambda^2$ and $b=1+\mu^2$)
\begin{equation}\label{hp-scale}
Z^{hp}_n(a)
  =\lim_{w\rightarrow\infty}Z_{n,w}(a,b)\sim
  \begin{cases}
  \left[\frac{2 (1+\lambda^2)}{(1-\lambda^2)^2}\sqrt{\frac{2}{\pi}}\,\right]\;
  \frac{2^n}{n^{3/2}}& 0\leq a <2\; (0\leq\lambda<1),\\[1ex]
  \left[ \sqrt{\frac{2}{\pi}}\,\right] \; \frac{2^n}{n^{1/2}}& a=2\,
  (\lambda=1),\\[1ex]
  \left[\frac{(\lambda^2-1)}{\lambda^2}\right] \,
  \left(\lambda+\lambda^{-1}\right)^n  &  a>2\, (\lambda>1),
  \end{cases}
\end{equation}
as $n \to \infty$. Note that the scaling is independent of $b$, as it should be
for $w>n$. For any finite $w$ we expect that
\begin{equation}\label{finite-w-scale}
Z_{n,w}(a,b) \sim B_w(a,b) \; \mu_w(a,b)^n
\end{equation}
where as $w\rightarrow \infty$ we have
\begin{equation}\label{inf-slit-mu}
\mu_{w}(a,b)\rightarrow\begin{cases}
  2                                & 0\leq a,b \leq 2,\\
\left(\lambda+\lambda^{-1}\right)  & a\geq b \text{ and } a>2,\\
\left(\mu+\mu^{-1}\right)  & a<b \text{ and } b>2.
\end{cases}\end{equation}

If we denote the right-hand side of equation~(\ref{hp-scale}) as
$S^{hp}(n)$ for each value of $a$ then a canonical general scaling Ansatz is
\begin{equation}\label{scaling}
Z_{n,w}(a,b) \sim S^{hp}(n)\;
f_{\text{phase}}\left(\frac{n^{\nu_{\perp}}}{w}\right).
\end{equation}
The scaling function $f_{\text{phase}}$ depends on which phase
or phase boundary of the infinite slit the values of $a$ and $b$
correspond (so as to match with the infinite slit phases).The
exponent $\nu_{\perp}$, on the other hand, should depend on the phase of the
half-plane problem.

We immediately note that there is something unusual here and
that while $\nu_\perp$ is expected to be $1/2$ for $0\leq a\leq 2$ it is
expected to be $0$ for $a>2$\;!  In Martin \emph{et al.\ }\cite{martin2007a-:a}
it was proposed that a scaling theory like that in equation~(\ref{scaling}) 
(see equations~(3.5),~(3.6) and~(3.7) of that paper)
should only hold in the desorbed region of the infinite slit or on its
boundaries. That is, we should expect a scaling theory only
when $0\leq a, b\leq 2$ --- this is precisely when $\nu_\perp =1/2 >0$.
This is what we have found for the directed walk case.

Now, to \emph{match} the scaling in equation~(\ref{hp-scale}) (half-plane
matching) one would require that
\begin{equation}\label{match-hp}
 f_{\text{phase}}(x) \sim  1 \text{ as } x \rightarrow 0\,.
\end{equation}
This is a similar requirement to that of equation (3.6) of Martin \emph{et al.\ }\cite{martin2007a-:a}.
On the other hand the scaling of $f_{\text{phase}}$ as $x\rightarrow \infty$ needs to be
considered in order to match the scaling of the \emph{infinite slit}
(equations~(\ref{finite-w-scale}) and~(\ref{inf-slit-mu})).

Our results show that the  partition function $Z_{n,w}(a,b)$ can be
found as two parts $Z^{(e)}$ and $Z^{(s)}$ as given in
equations~(\ref{part-e-scale}) and~(\ref{part-s-scale})
respectively. Importantly, for $0\leq a,b\leq 2$ we have $Z^{(e)}=0$ and
the form~(\ref{part-s-scale}) for $Z^{(s)}$ is precisely that of~(\ref{scaling})
with the half-plane matching behaviour~(\ref{match-hp}) obeyed
by~(\ref{scale-fns}). The functions~(\ref{scale-fns}) also obey the required
\emph{infinite slit} matching behaviour: that is,
\begin{equation}\label{match-inf-slit}
 f_{\text{phase}}(x) \propto \begin{cases}
\ds
x
&\quad\text{for $a=b=2$}\\[2ex]
\ds
x^3 e^{-\pi^2x^2/8}
&\quad\text{for $a<2$ and $b=2$,}\\[2ex]
\ds
x e^{-\pi^2x^2/8}
&\quad\text{for $a=2$ and $b<2$, and}\\[2ex]
\ds
x^3e^{-\pi^2x^2/2}
&\quad\text{for $a<2$ and $b<2$}
\end{cases}
\end{equation}
as $x \rightarrow \infty$. This large $x$ behaviour allows for the
recovery of~(\ref{finite-w-scale}) with~(\ref{inf-slit-mu}). They
also give us the first correction to the free energy as a function of
$w$ (see equations (7.3) and~(7.6) of Brak \emph{et al.\
}\cite{brak2005a-:a}). From equation~(3.7) of Martin \emph{et al.\
}\cite{martin2007a-:a} we find that the above results also agree with
that generic prediction.

For any value of $(a,b)$ such that either $a>2$ and/or $b>2$,
$Z^{(e)}\neq0$ and furthermore $Z^{(e)} >> Z^{(s)}$ as $n\rightarrow
\infty$. As far as we can calculate, $Z^{(e)}$ cannot be written in
the scaling function form~(\ref{scaling}). We note that the scaling
form of $Z^{(s)}$ acts as a correction to scaling to that of $Z^{(e)}$
which allows for a \emph{matching} of the total form $Z^{(e)}+Z^{(s)}$
with the half-plane and infinite slit limits. Since the dominant part
of the form of $Z^{(e)}$ in~(\ref{part-e-scale}) is that
of~(\ref{finite-w-scale}) with the connective constant as
in~(\ref{inf-slit-mu}) the infinite slit behaviour is clearly adhered
to (although the coefficient $B_w(a,b)$ may go to zero in this limit).

Consider equation~(\ref{part-e-scale}). For $b>2$ and $a\leq 2$
(region ${\cal T}^{(1)}$) $Z^{(e)}\sim \mu^{-2w}\rightarrow 0$ as
$w\rightarrow \infty$ ($n$ fixed and large) so for the half-plane one
is left with $Z^{(s)}$ with $x\rightarrow 0$.  For $a>2$ and $b\leq 2$
(region ${\cal B}^{(1)}$) the scaling form in~(\ref{part-e-scale})
coincides with that of~(\ref{hp-scale}) for $a>2$ so that half-plane
limit is simple. For $a=b >2$ (line ${\cal S_{TB}}$) this is also true. For $a>2$ and $b>2$
with $a\neq b$ (regions ${\cal T}^{(2)}$ and ${\cal B}^{(2)}$) there are
two parts of the scaling form of $Z^{(e)}$ with one part going to zero
for $w\rightarrow \infty$ leaving the appropriate scaling
corresponding to the half-plane.

The above shows how complicated, though mathematically complete, the
scaling picture can be for this problem. To summarise, we have
calculated the scaling for large widths and large lengths of directed
walks confined between two walls that interact with the walk. We
explicitly demonstrate that the conjectured scaling theory
\cite{martin2007a-:a} for polymers confined in such a manner holds
exactly for this model. This theory holds when the polymer is in a
desorbed state, or on the boundaries of this region in the parameter
space, that is critically adsorbing.

 \section*{Acknowledgments}
 Financial support from NSERC of Canada, the Australian Research Council and
the Centre of Excellence for Mathematics and Statistics of Complex Systems is
 gratefully acknowledged by the authors.

\end{document}